\documentclass[11pt]{article}
\usepackage{fa2023}
\usepackage{amsmath}
\usepackage{cite}
\usepackage{tabularx}
\usepackage{url}
\usepackage{graphicx}
\usepackage{color}
\usepackage{siunitx}
\usepackage[utf8]{inputenc}
\usepackage{amsfonts}    
\usepackage{IEEEtrantools}
\title{Circumvent spherical Bessel function nulls for open sphere 
microphone arrays with physics informed neural network}
\multauthor
{Fei Ma 
\hspace{1cm} 
Thushara D. Abhayapala
\hspace{1cm} Prasanga N. Samarasinghe
} 
{College of Engineering, Computing $\&$ Cybernetics,  
Australian National University, Australia  \\
\correspondingauthor{fei.ma@anu.edu.au}}
\begin{document}
\maketitle
\begin{abstract}
Open sphere microphone arrays (OSMAs) are simple to design and do not introduce 
scattering fields, and thus can be advantageous than other arrays for  
implementing spatial acoustic algorithms under spherical model decomposition.
However, an OSMA suffers from spherical Bessel function nulls which 
make it hard to obtain some sound field coefficients at certain frequencies. 
This paper proposes to assist an OSMA for sound field analysis with 
physics informed neural network (PINN).
A PINN models the measurement of an OSMA and predicts the sound field on 
another sphere whose radius is different from that of the OSMA. 
Thanks to the fact that spherical Bessel function nulls vary with radius, 
the sound field coefficients which are hard to obtain based on the OSMA 
measurement directly can be obtained based on the prediction. 
Simulations confirm the effectiveness of this approach and compare 
it with the rigid sphere approach. 
\end{abstract}

\section{Introduction}
The products of the spherical harmonics (SHs) and the spherical Bessel functions 
(or the spherical Hankel functions) form the spherical nodes~\cite{williams2000fourier}, 
a complete and orthogonal function set for the Helmholtz equation, the governing partial 
differential equation (PDE) of acoustic wave propagation. 
The SH decomposition of a sound field  (the 
angular dependent SHs, the radial dependent spherical Bessel functions, and the 
frequency dependent sound field coefficients) greatly facilitates its analysis 
and manipulation~\cite{williams2000fourier,thushara_near_1999,Rafaely2015}.
Thus, spherical modal decomposition has become popular in many diverse spatial 
acoustic applications, such as spatial active noise control~\cite{Wen_2018,ma2018active,Fei_2020},
beamforming~\cite{5745011,Rafaely2015,huang2018insights}, 
and direction of arrival estimation~\cite{moore2016direction, 
hafezi2017augmented,jo2019parametric}.

Due to their simplicity, open sphere microphone arrays 
(OSMAs) are intuitively chosen 
for implementing the spherical modal decomposition~\cite{5745011}.  
However, the spherical Bessel function nulls make it hard to obtain some order
of the sound field coefficients at certain frequencies with an OSMA.
We can mitigate this problem through arranging microphones 
on a rigid sphere~\cite{5744968}, inside a spherical shell, or using vector 
sensors on an open sphere~\cite{Rafaely2015,Ma2018,huang_flexible}. 
However, those approaches will unavoidably introduce scattering fields,  
request more microphones, and significantly increase the cost, respectively. 

In this paper, we propose to circumvent the problem of spherical Bessel function nulls 
for an OSMA with the help of physics informed neural network (PINN)~\cite{raissi2019physics,cuomo2022scientific,karniadakis2021physics}, a
neural work which incorporates physical knowledge into its architecture and training. 
We model the measurement of an OSMA with a PINN, and then use it to predict the sound 
field on another sphere whose radius is different from that of the OSMA.  
Thanks to the fact that the spherical Bessel function nulls vary with radius, we can 
obtain the sound field coefficients which are difficult to obtain with the OSMA 
measurement based on the predicted sound field.
The effectiveness of this approach is confirmed by simulations and compared with the rigid 
sphere approach. 

\section{Problem formulation}
\begin{figure}[t]
\centerline{\framebox{
\includegraphics[width=7cm]{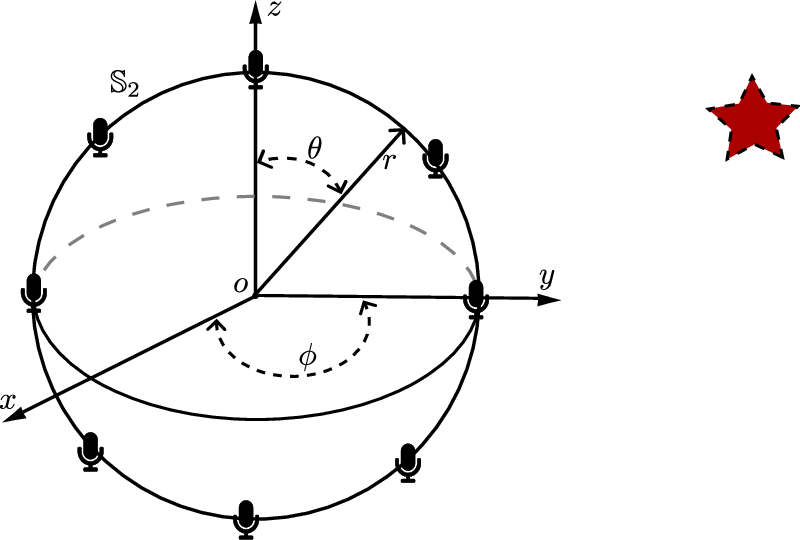}}}
\caption{A microphone array on an open sphere and some sound sources $\star$.}
\label{fig:bf}
\end{figure}
We consider the set up shown in Fig.~\ref{fig:bf}, where there are $Q$ omni-directional 
pressure microphones on an open sphere $\mathbb{S}_2$ of radius $r_a$ and some sound sources. 
The Cartesian coordinates and the spherical coordinates of a point 
with respect to an origin are denoted as $O$ as $(x,y,z)$  and $(r,\theta,\phi)$, respectively 
\cite{Rafaely2015}. 
One would like to reconstruct the sound field around the sphere or 
locate the sound sources based on the OSMA measurement.

The tasks could be approached with SH decomposition.  
We decompose the sound pressure at microphone position 
$\{(r_a,\theta_q,\phi_q)\}_{q=1}^{Q}$ onto SHs as \cite{williams2000fourier}
\begin{IEEEeqnarray}{rcl}
\label{eq:shd}
{P}(\omega,r_a,\theta_q,\phi_q)
&\approx&\sum_{u=0}^{U}\sum_{v=-u}^{u}\mathsf{P}_{u,v}(\omega,r_a)Y_{u,v}(\theta_q,\phi_q) \nonumber\\
&=&\sum_{u=0}^{U}\sum_{v=-u}^{u}\mathsf{K}_{u,v}(\omega)
{j}_{u}(\omega{}r_a/s) 
\nonumber\\
&&\times 
Y_{u,v}(\theta_q,\phi_q),	
\end{IEEEeqnarray}
where $\omega=2\pi{f}$ is the angular frequency ($f$ is the frequency),  $s$ is the speed 
of sound  propagation, $U=\lceil{2\pi{f}r_a/s}\rceil$ is the up-order of the SHs that are needed 
to represent the sound pressure accurately~\cite{Thusharahigh} 
($\lceil{\cdot}\rceil$ is the ceiling operation),.
$\mathsf{P}_{u,v}(\omega,r_a)$ are the pressure field coefficients, 
$\mathsf{K}_{u,v}(\omega)$ are the sound field coefficients~\cite{williams2000fourier}, 
$j_{u}(\cdot)$ is the spherical Bessel function of the first kind of 
order $u$, $Y_{u,v}(\theta,\phi)$ is the SH of order $u$ and degree 
$v$~\cite{williams2000fourier} at is evaluated at $(\theta,\phi)$.

\begin{figure}[t]
\centerline{\framebox{
\includegraphics[width=8cm]{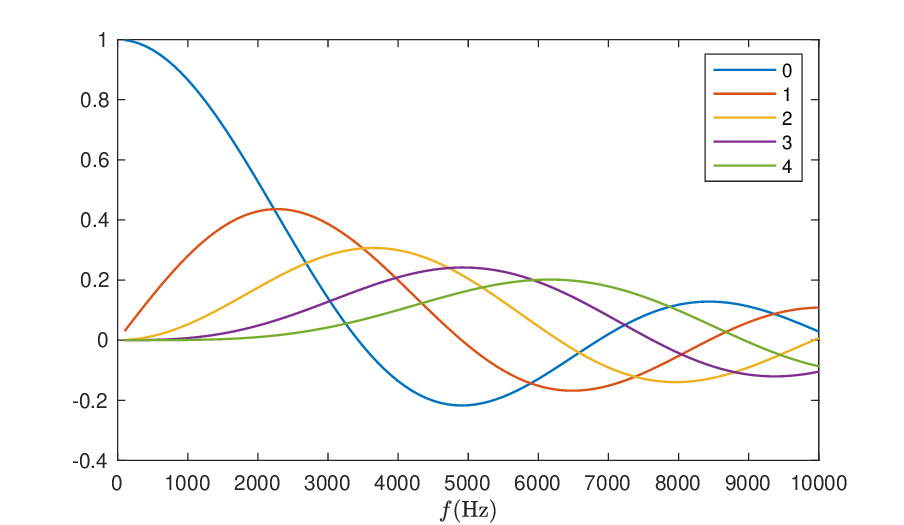}}}
\caption{The spherical Bessel function $j_u(2\pi{f}r_a/s)$  
as a function of frequency, $r_a=0.05$ m, $s=343$ m/s, $u=0,1,2,3,4$.}
\label{fig:jnx}
\end{figure}

The sound field coefficients $\mathsf{K}_{u,v}(\omega)$ characterize the sound 
sources and allow us to reconstruct the sound field or to locate the sound 
sources~\cite{Rafaely2015}. 
To obtain the sound field coefficients, we first estimate the pressure field 
coefficients through \cite{Rafaely2015}
\begin{IEEEeqnarray}{rcl}
\label{eq:puv}
\hat{\mathsf{P}}_{u,v}(\omega,r_a)=\sum_{q=1}^{Q} P(\omega,r_a,\theta_q,\phi_q)Y_{u,v}(\theta_q,\phi_q)\gamma_q,
\end{IEEEeqnarray}
where $\{\gamma_q\}_{q=1}^{Q}$ are the sampling weights, and then 
 estimate the sound field coefficients through
\begin{IEEEeqnarray}{rcl}
\label{eq:kuv}
\hat{\mathsf{K}}_{u,v}(\omega)
&=&{\hat{\mathsf{P}}_{u,v}(\omega,r_a)}/{{j}_{u}(\omega{}r_a/s)}. 
\end{IEEEeqnarray}
The problem with \eqref{eq:kuv} is the spherical Bessel function 
${j}_{u}(\cdot)$ nulls \cite{Rafaely2015,Ma2018,huang_flexible}. 
Figure~\ref{fig:jnx} presents $j_u(2\pi{f}r_a/s)$ with $r_a=0.05$ m, 
$s=343$ m/s, $u=0, 1, 2, 3, 4$.
We can see that $j_0(2\pi{f}r_a/s)=0$ for $f=3430$ Hz,   
and $j_1(2\pi{f}r_a/s)=0$ for $f=4905$ Hz. 
This makes it  difficult to estimate the sound field coefficients of order 0, 
$\mathsf{K}_{0,0}(\omega)$, and order 1, $\mathsf{K}_{1,v}(\omega)$,
at frequency 3430 Hz and 4905 Hz with an OSMA array of radius $r_a=0.05$ m.  

In this paper, 
we aim to circumvent the problem of spherical 
Bessel function nulls for an OSMA. 
\section{PINN Assisted OSMA}
In this section, we propose a PINN method to assist an OSMA for sound field analysis.
To simplify the calculation of the Laplacian,   we express acoustic quantities in 
Cartesian coordinates. 

The key idea is to exploit the fact that spherical Bessel function nulls vary with 
radius. 
The spherical Bessel function $j_u(\cdot)$ is a function of both frequency $f$ and 
radius $r$, and thus that $j_u(2\pi{f}r_b/s)\neq0$ if $r_b{}\neq{}r_a$ 
and $j_u(2\pi{f}r_a/s)=0$.  
Thus that we can obtain the  sound field coefficients which are difficult 
to obtain with  an OSMA of radius $r_a$ 
based on the sound field on another sphere of radius $r_b$.
An OSMA of radius $r_a$ can not  measure the sound field on 
another sphere of radius $r_b$ directly, but we can build a 
PINN~\cite{raissi2019physics,cuomo2022scientific,karniadakis2021physics} to 
predict the sound  field on the other sphere based on the measurement of the OSMA.

We build up a $L$ layer $N$ node (on each layer) full connected feedforward neural 
network \cite{raissi2019physics} whose inputs are Cartesian coordinates $(x,y,z)$ 
and output is the sound field estimation $\hat{P}_{\mathrm{PI}}(\omega,x,y,z)$,
and update the trainable parameters of the network by minimizing the following cost 
function
\begin{IEEEeqnarray}{rcl}
\label{eq:cost}
&&\mathfrak{L}= 
\underbrace{
\frac{1}{Q}\sum_{q=1}^{Q}\|
P(\omega,x_q,y_q,z_q)-
\hat{P}_{\mathrm{PI}}(\omega,x_q,y_q,z_q)
\|_2^2
}_{\mathfrak{L}_{\mathrm{data}}}
\nonumber\\
&&+
\underbrace{
\frac{1}{A}
\sum_{a=1}^{A}
\| 
\frac{
\nabla{}\hat{P}_{\mathrm{PI}}(\omega,x_a,y_a,z_a) 
}
{(w/s)^2}
+\hat{P}_{\mathrm{PI}}(\omega,x_a,y_a,z_a) \|_2^2
}_{\mathfrak{L}_{\mathrm{PDE}}}, \quad\;
\end{IEEEeqnarray}
where $\|\cdot\|_2$ is the 2-norm, 
$ \nabla\equiv \frac{\partial^2}{\partial{}x^2} 
+ \frac{\partial^2}{\partial{}y^2}    
+ \frac{\partial^2}{\partial{}z^2}$ 
is the Laplacian. 
The data loss $\mathfrak{L}_{\mathrm{data}}$ makes the network output to approximate 
the OSMA measurement $\{P(\omega,x_q,y_q,z_q)\}_{q=1}^{Q}$
where $(x_q, y_q, z_q)_{q=1}^{Q}$ correspond to $(r_a,\theta_q,\phi_q)_{q=1}^{Q}$.
The PDE loss  $\mathfrak{L}_{\mathrm{PDE}}$ informs the network output to conform 
with the Helmholtz equation on the measurement sphere of radius $r_a$, where 
$\{(x_a,y_a,z_a)\}_{a=1}^{A}$ are uniformly arranged sampling points on the sphere. 

To obtain the sound field coefficients, we first train the PINN and use it to estimate 
the pressure $\hat{P}_\mathrm{PI}(\omega,x_d,y_d,z_d)$ (which are equal to
$\hat{P}_\mathrm{PI}(\omega,r_b,\theta_d,\phi_d)$) on a sphere of radius $r_b$.
Next we estimate the pressure field coefficients $\hat{\mathsf{P}}_{u,v}(\omega,r_b)$ 
\begin{IEEEeqnarray}{rcl}
\label{eq:puvb}
\hat{\mathsf{P}}_{u,v}(\omega,r_b)=\sum_{d=1}^{D} 
\hat{P}_\mathrm{PI}(\omega,r_b,\theta_d,\phi_d)Y_{u,v}(\theta_d,\phi_d)\gamma_d,
\end{IEEEeqnarray}
where $\{\gamma_d\}_{d=1}^{D}$ are the sampling weights~\cite{Rafaely2015}. 
We further estimate the sound field coefficients through
\begin{IEEEeqnarray}{rcl}
\label{eq:kuvb}
\hat{\mathsf{K}}_{u,v}(\omega)
&=& {\hat{\mathsf{P}}_{u,v}(\omega,r_b)}/{{j}_{u}(\omega{}r_b/s)}. 
\end{IEEEeqnarray}
In summary, for spatial acoustics with an OSMA, we can estimate the sound field 
coefficients through \eqref{eq:kuv} when $j_u(kr_a)\neq0$ and through 
\eqref{eq:cost}, \eqref{eq:puvb}, \eqref{eq:kuvb} when $j_u(kr_a)=0$.
In this way, the problem of spherical Bessel function nulls is circumvented.

Note that the spherical Bessel function nulls is a problem under the spherical modal
decomposition, but it is not a problem with the PINN. This is the fundamental 
fact that make the PINN assisted OSMA sound field analysis possible. 

\section{Simulation}
In this section, we use a sound field reconstruction task to demonstrate the performance 
of the PINN assisted OSMA and compare it with the rigid sphere approach.

We consider the setup  shown in Fig.~\ref{fig:bf}. 
There is a radius $r_a=0.05$ m OSMA with 36  uniformly arranged omni-directional 
pressure microphones on it. 
There is a sound source located at $(0.5\; \mathrm{m, 0.5\;\mathrm{m}}, 0.75\; \mathrm{m})$. 
The sound source generates a unit amplitude sinusoidal signal at $f=3430$ Hz.
In the case, the up-order of SHs needed to represent the sound field is 
$U=\lceil{2\pi\times3430\times0.05/343}\rceil$ = 4~\cite{Thusharahigh}.
The transfer functions between the sound source and the microphones 
are simulated  based on the Green's function~\cite{williams2000fourier}.
The aim is to reconstruct the sound field on a smaller sphere of radius $r_c=0.04$ m. 

Three approaches for sound field reconstruction are considered. 
The first is the OSMA approach based on the spherical modal decomposition. 
For this approach, we estimate the sound field coefficients 
$\{\hat{\mathsf{K}}_{u,v}(\omega)\}_{u=1}^{4}$ through \eqref{eq:kuv} and 
reconstruct the sound field on the smaller sphere by
\begin{IEEEeqnarray}{rcl}
\label{eq:shd}
\hat{P}_{\mathrm{SH}}(\omega,r_c,\theta,\phi)
&\approx&\sum_{u=1}^{U}\sum_{v=-u}^{u}\hat{\mathsf{K}}_{u,v}(\omega)
{j}_{u}(\omega{}r_c/s) 
Y_{u,v}(\theta,\phi),\quad \;
\end{IEEEeqnarray}
because $\hat{\mathsf{K}}_{0,0}(\omega)$ is not obtainable. 

\begin{figure}[t]
\centerline{\framebox{
\includegraphics[trim={0.4cm 1.6cm 1.20cm 1.2cm},clip,width=6cm,height=12cm]{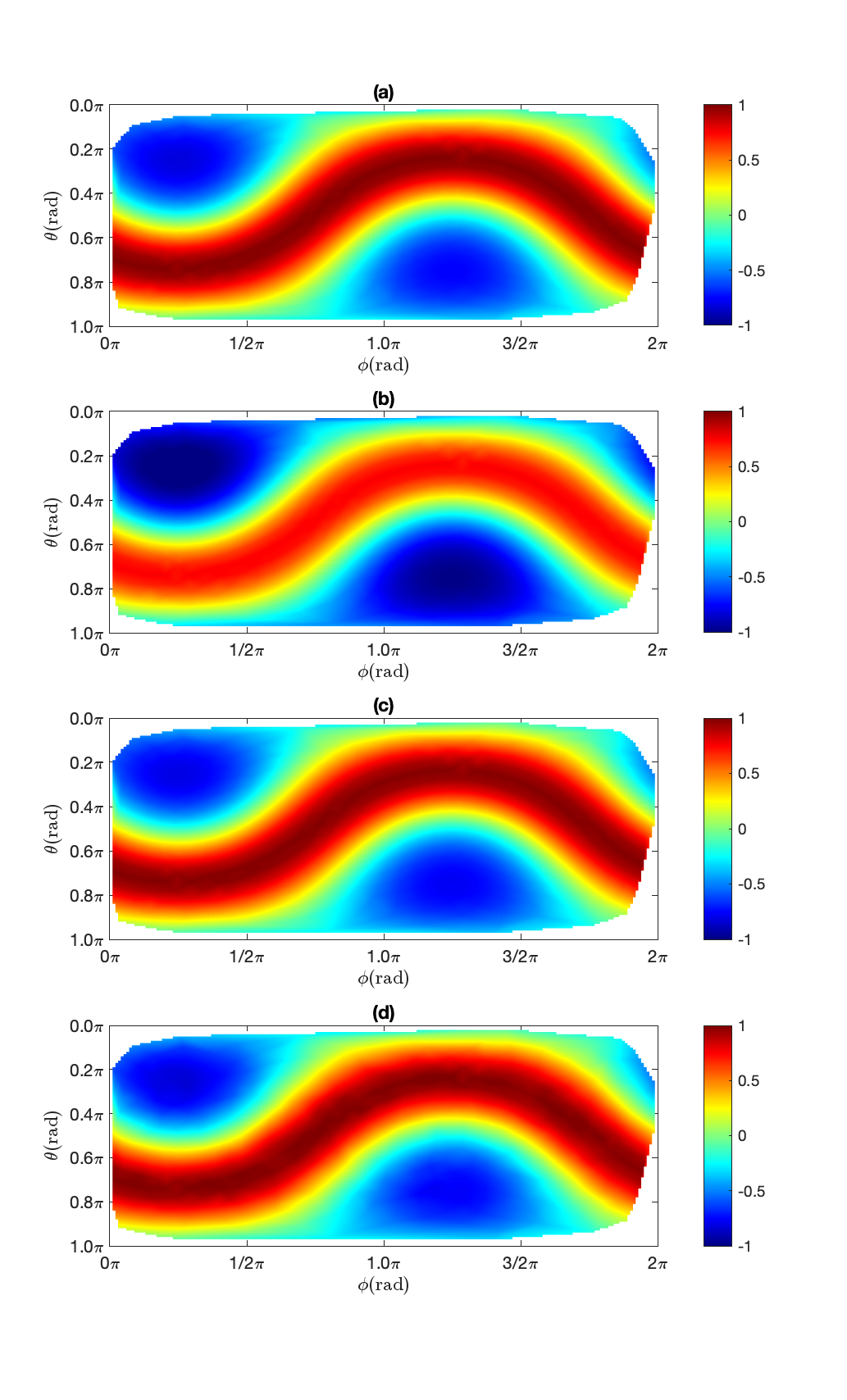}}}
\caption{Sound field at $f=3430$ Hz : (a) the ground truth, 
(b) OSMA reconstruction, (c) PINN assisted OSMA reconstruction, and (d) the rigid sphere reconstruction.}
\label{fig:field}
\end{figure}

\begin{figure}[t]
	\centerline{\framebox{
			\includegraphics[trim={0.4cm 1.6cm 1.20cm 1.2cm},clip,width=6cm,height=12cm]{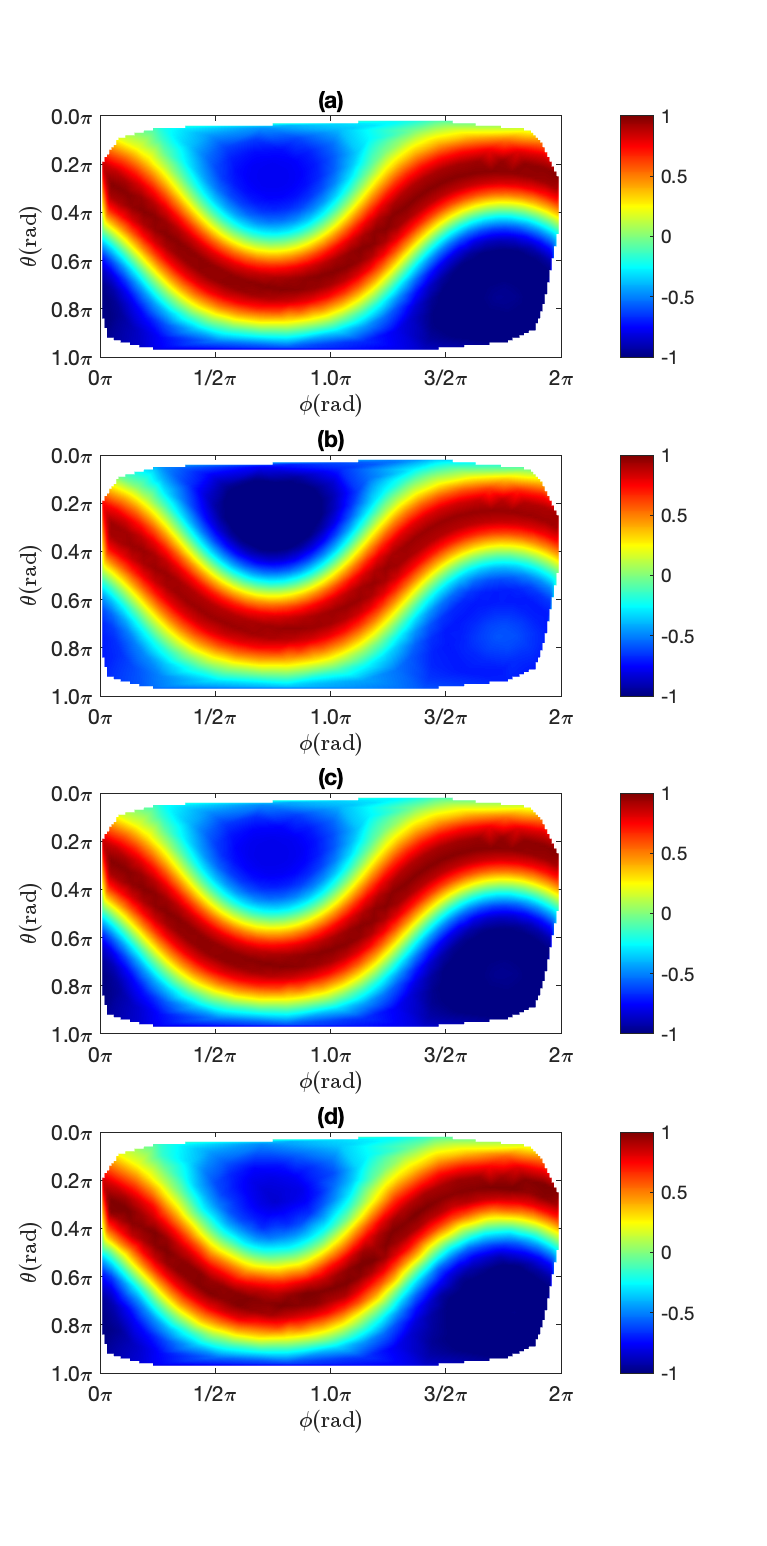}}}
	\caption{Sound field at $f=4905$ Hz : (a) the ground truth, 
		(b) OSMA reconstruction, (c) PINN assisted OSMA reconstruction, and (d) the rigid sphere reconstruction.}
\label{fig:4095}
\end{figure}

\begin{figure}[t]
\centerline{\framebox{
\includegraphics[width=7.5cm]{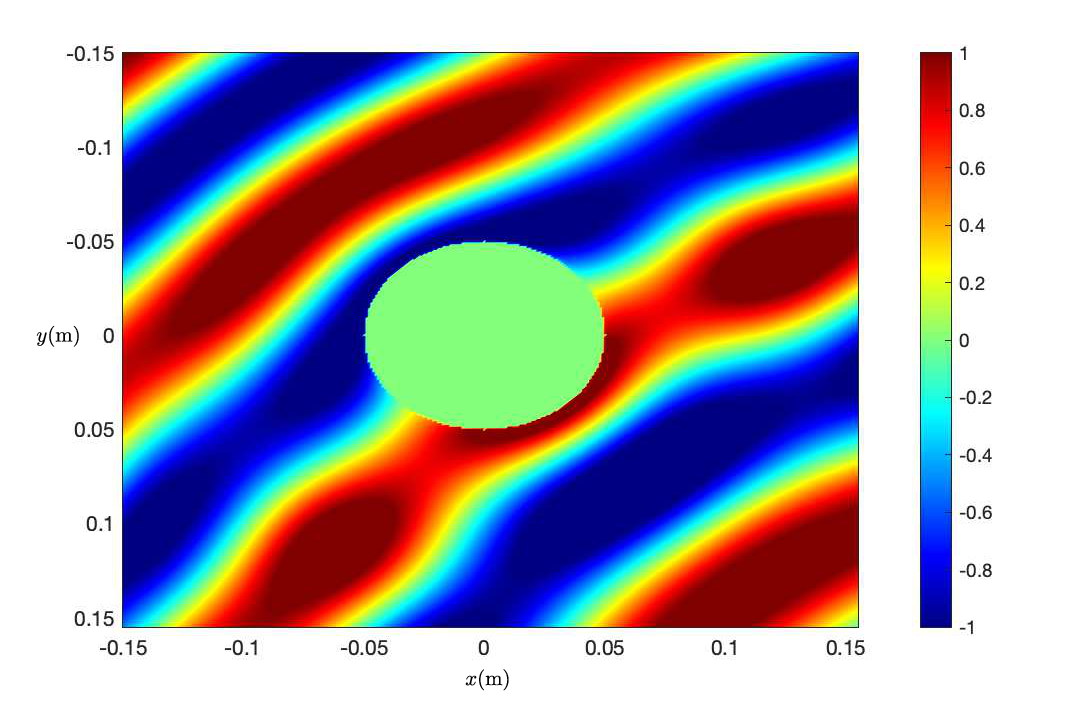}}}
\caption{The scattering field around a rigid sphere at 3430 Hz.}
\label{fig:scatter}
\end{figure}

\begin{figure}[t]
\centerline{\framebox{
\includegraphics[trim={0.75cm 0.0cm 1.cm 0.4cm},clip,width=7.5cm]{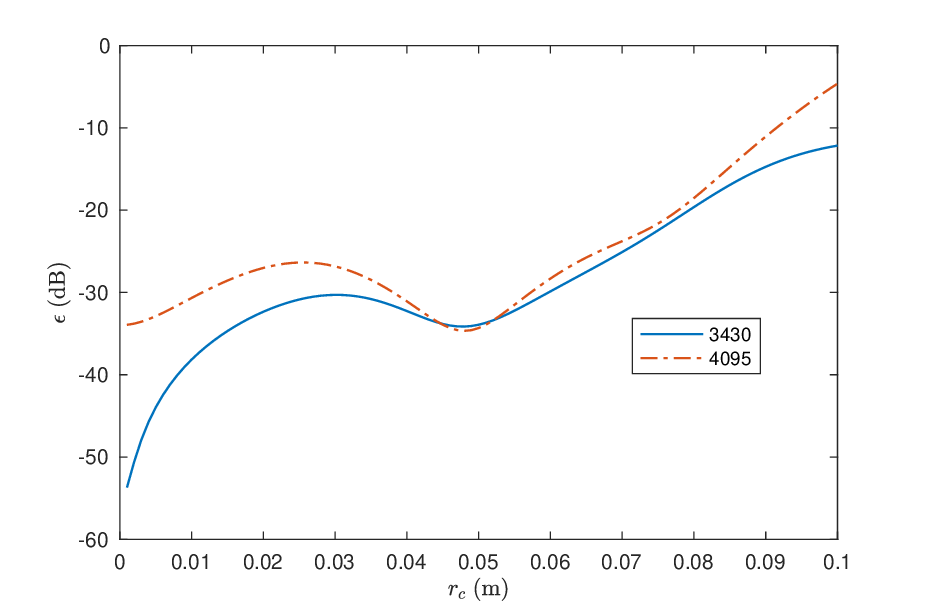}}}
\caption{Sound field reconstruction error $\epsilon$ of the pure PINN method as a 
function of the reconstruction sphere radius $r$ at 3430 Hz and 4095 Hz.}
\label{fig:freq}
\end{figure}

The second one is the PINN assisted OSMA method. For this method, we build up a $L=3$ layer 
and $N=3$ node PINN, with the activation function being $\tanh$, and 
initialize the trainable parameters with the Xavier initialization~\cite{glorot2010understanding}.
PINN is trained for 10$^8$ epochs with a learning rate of 10$^{-5}$ 
using the ADAM optimizer. 
The data loss $\mathfrak{L}_\mathrm{data}$ is evaluated with respect to the 
36 microphone measurements, and the PDE losses  $\mathfrak{L}_{\mathrm{PDE}}$  
with respect to the Cartesian coordinates of 500 uniformly arranged sampling 
points on the sphere of radius $r_a=0.05$ m.
We first estimate the sound field coefficients $\{\hat{\mathsf{K}}_{u,v}(\omega)\}_{u=1}^{4}$ 
through \eqref{eq:kuv}, $\hat{\mathsf{K}}_{0,0}(\omega)$ through \eqref{eq:cost},
\eqref{eq:puvb} with \eqref{eq:kuvb} with $r_b=0.048$ m, 
and next reconstruct the sound field on the smaller sphere similar as \eqref{eq:shd} but with 
$\hat{K}_{0,0}(\omega)j_u(\omega{r_c}/s)Y_{0,0}(\theta,\phi)$ included. 

The third one is the rigid sphere approach. 
The OSMA in Fig.~\ref{fig:bf} is replaced with a rigid sphere of the same radius, and the 
rest of simulation setting is the same. 
we reconstruct the sound field on the sphere of radius $r_c$ as 
\begin{IEEEeqnarray}{rcl}
P(\omega,r_c,\theta,\phi)&\approx&\sum_{u=0}^{U} G_u(\omega,r_c,r_a) \nonumber\\
&&\times
\sum_{v=-u}^u\hat{\mathsf{P}}_{u,v}(\omega,r_a) Y_{u,v}(\theta, \phi),
\label{eq:1}
\end{IEEEeqnarray}
where 
the pressure field coefficients $\hat{\mathsf{P}}_{u,v}(\omega,r_a)$ are obtained similar to \eqref{eq:puv}, 
$G_u(\omega,r_c,r_a)$ is the radial translator~\cite{Rafaely2015} 
\begin{IEEEeqnarray}{rcl}
G_u(\omega,r_c,r_a)&=&
\frac
{h_u^{\prime}({\omega r_a}/{s}) j_u(\omega{r_c}/s)}
{j_u({\omega r_a}/{s}) h_u^{\prime}({\omega r_a}/{s}) - j_u^{\prime}({\omega r_a}/{s}) h_n({\omega r_a}/{s})}
, 
\nonumber\\
\end{IEEEeqnarray}
$j_u(\cdot)$ and $h_u(\cdot)$ are the spherical Bessel function of the first kind and the spherical 
Hankel function of the second kind, respectively, and $j_u^{\prime}(\cdot)$ and $h_u^{\prime}(\cdot)$
are corresponding derivatives with respect to argument.

We denote the reconstruction 
error as 
\begin{IEEEeqnarray}{rcl}
\epsilon=\frac{
\sum_{d=1}^{100}
||
{P}(\omega,r_c,\theta_d,\phi_d)-
\hat{P}(\omega,r_c,\theta_d,\phi_d)
||_2^2
}{
\sum_{d=1}^{100}
||{P}(\omega,r_c,\theta_d,\phi_d)||_2^2
},\quad 
\end{IEEEeqnarray}
where ${P}(\omega,r_c,\theta_d,\phi_d)$ and $\hat{P}(\omega,r_c,\theta_d,\phi_d)$
are the true pressure and its reconstruction at 100 uniformly selected sampling positions  $(\theta_d,\phi_d)_{d=1}^{100}$.

Real part of the ground truth and its reconstructions 
by three methods is shown in Fig.~\ref{fig:field}. 
Comparing Fig.~\ref{fig:field} (b) and (a), we can see that  the sound field 
component $K_{0,0}(\omega)j_u(\omega{r_c}/s)Y_{0,0}(\theta,\phi)$ missing the OSMA approach is unable to 
accurately reconstruct the ground truth. 
Comparing Fig.~\ref{fig:field} (c), (d) and (a), we can see that with 
the PINN assisted OSMA approach and the rigid sphere approach 
are able to accurately reconstruct the sound field. 
The reconstruction errors of the OSMA approach, the PINN assisted OSMA approach, 
and the rigid sphere approach are -8.5 dB, -28.4 dB and -29.3 dB, respectively. 
The simulation results of three approaches for reconstructing the imaginary part 
of the ground truth are similar to Fig.~\ref{fig:field}, and thus are not shown for brevity.

The simulations is repeated at $f=4095$ Hz. We arrange the sound source at $(0.5\; \mathrm{m}, -0.5\;\mathrm{m}, -0.75\;\mathrm{m})$ 
and the rest of simulation settings are the same as the $f=3430$ Hz case.     
Real part of the ground truth and its reconstructions 
by three methods is shown in Fig.~\ref{fig:4095}. 
The reconstruction errors of the OSMA approach, the PINN assisted OSMA approach, 
and the rigid sphere approach are -10.2 dB, -31.6 dB and -32.3 dB, respectively.

The PINN-assisted OSMA approach performs comparably to the rigid sphere approach.
Nonetheless, the rigid sphere approach has a drawback: the scattering effect. 
In Fig.~\ref{fig:scatter}, we present the scattering field around the rigid 
sphere at 3430 Hz. For nearfield applications, where the rigid sphere is placed 
close to some object, the scattering field will further undergo multiple scatterings 
and is highly undesirable~\cite{colton1998inverse}. 
The PINN assisted OSMA approach, on the other hand, does not suffer from the scattering problems.

Figure~\ref{fig:freq}  presents the sound field reconstruction error $\epsilon$
of a pure PINN method as a function of reconstruction sphere radius $r_c$ at 3430 Hz and 4095 Hz.
In this case, the sound field is reconstructed based on the PINN prediction directly and only, 
and does not go through the SH decomposition process. 
From Fig.~\ref{fig:freq}, we can see that the reconstruction error $\epsilon$ is small when the 
reconstruction sphere radius is close  to the array radius, or $r_c\approx{}r_a=0.05$ m, and increases 
when they are not close $r_c\not\approx{}r_a=0.05$ m. 
It is interesting that the reconstruction error  $\epsilon$ decreases again when the reconstruction 
sphere radius is  small, or $r_c<0.025$ m. From  a  spherical modal decomposition point of view,
this can be explained by the fact that when the reconstruction sphere radius is  small less 
number of SH coefficients are needed to describe the sound field.

\section{Conclusion}
In this paper, we proposed to assist an OSMA for sound field analysis with the PINN. 
Under SH decomposition, 
the OSMA suffers from the spherical Bessel function nulls and is unable to obtain
some orders of sound field coefficients at certain frequencies. We use a PINN to predict 
the sound field on a sphere whose radius is different from that of the OSMA, and 
obtain those order of sound field coefficients based on the PINN prediction. 
The performance of this approach is comparable with  the rigid sphere approach  and  
does not introduce the scattering field. 


\section{Acknowledgments}
We thank  Hanwen Bi  from Australian National University for reviewing the simulation code, which has lead to 
significant performance improvement of the simulation results. 

\bibliography{fa2023}
\end{document}